# Ensemble-based Multi-Filter Feature Selection Method for DDoS Detection in Cloud Computing

Opeyemi Osanaiye[1,2], Kim-Kwang Raymond Choo[2*], Ali Dehghantanha[3], Zheng Xu[4,5], Mqhele Dlodlo[1]

[1]Department of Electrical Engineering, University of Cape Town, South Africa

[2]Information Assurance Research Group, University of South Australia, South Australia 5095, Australia

[3]School of Computing, Engineering and Technology, University of Salford, Manchester

[4]Tsinghua University, Beijing, China

[5]Third Research Institute of the Ministry of Public Security, Shanghai, China

opyosa001@myuct.ac.za, raymond.choo@fulbrightmail.org [Corresponding Author],

A.Dehghantanha@salford.ac.uk, Xuzheng@shu.edu.cn, mqhele.dlodlo@uct.ac.za

**Abstract**

Increasing interest in the adoption of cloud computing has exposed it to cyber-attacks. One of such is distributed denial of service (DDoS) attack that targets cloud's bandwidth, services and resources to make it unavailable to both the cloud providers and users. Due to the magnitude of traffic that needs to be processed, data mining and machine learning classification algorithms have been proposed to classify normal packets from an anomaly. Feature selection has also been identified as a pre-processing phase in cloud DDoS attack defence that can potentially increase classification accuracy and reduce computational complexity by identifying important features from the original dataset, during supervised learning. In this work, we propose an ensemble-based multi-filter feature selection method that combines the output of four filter methods to achieve an optimum selection. An extensive experimental evaluation of our proposed method was performed using intrusion detection benchmark dataset, NSL-KDD and decision tree classifier. The result obtained shows that our proposed method effectively reduced the number of features from 41 to 13 and has a high detection rate and classification accuracy when compared to other classification techniques.

## 1. Introduction

The internet has been at the centre stage of recent advances in the ICT world. It is one of the major enabler of cloud computing used in providing services, resources and applications. Cloud computing has shifted the focus of both small and large organisations from the day-to-day running of their IT facilities by providing an on-demand, scalable and reliable computing resources [1]. Cloud can be deployed as either a private, public, community or hybrid and its service model can be broadly categorized into three; Software-as-a-service (SaaS), Platform-as-a-service (PaaS) and Infrastructure-as-a-service (IaaS) [2]. Even though the adoption of cloud computing offers numerous benefits, the open architecture of the internet has made it susceptible to cyber-attacks. DDoS attack has been reported in the literature to consume resources and deny legitimate cloud user's access [3]. In its simplest form, a handler recruits several vulnerable hosts, known as zombies, on the internet to direct a coordinated attack towards a pre-determined host. This attack has continued to increase both in size and sophistication, and extortion has been identified as one of the main motives behind the attack [4].

Proposed DDoS defense techniques are designed to classify packets as either legitimate or malicious and can be broadly categorized into signature-based and anomaly-based. Signature-based involves the use of well-known attack signatures in a knowledge database to determine the presence of an attack while anomaly based uses the collection of normal traffic behavioural pattern over a set time to classify subsequent patterns that deviate from the expected behaviour. Signature-based detection is very efficient in detecting well know attacks while anomaly detection can detect zero-day attacks. To counter the limitation associated with both approaches, a hybrid solution that uses both techniques has been proposed.

The internet in recent times has attracted a lot of users which has increased the amount of data that needs to be processed [37-38]. Feature selection is the pre-processing phase before classification that identifies important features of a dataset with the aim of improving prediction accuracy and reducing computational complexity. Current defense methods deal with huge amount of data that contains redundant or irrelevant features which result in excessive training and classification time [5]. Feature selection methods have been used in different areas of study such as statistical pattern recognition, machine learning and data mining for data reduction in other to achieve an improved performance and detect outliers.

Current feature selection methods can be categorized into three categories namey filter, wrapper and embedded methods. In filter methods, attributes are categorized according to the intrinsic information of the data and it is independent of the classification algorithm [6]. In filter method, features are accessed and ranked according to its inherent properties using simple measurement such as distance, dependency and information [7]. This makes it efficient when dealing with large data set as compared to wrapper methods that present a more precise result but are time- consuming [8]. Wrapper and embedded methods, on the other hand, are immersed in specific classification algorithm to determine the importance of a feature subset.

Recent studies have shown that combining feature selection methods would improve the performance of classifiers by identifying features that are weak as an individual but strong as a group[11], removing redundant features[8], and determining features that have a high correlation with the output class. Other methods have proposed a hybrid feature selection that combines both filter and wrapper. Filter feature selection represents a popular method that uses ranking and space search technique, therefore in this work, we present an Ensemble-based Multi-Filter Feature Selection (EMFFS) method that combines the output of Information Gain (IG), Gain Ratio, Chi-squared and reliefF to select important features. The aim of this work is to significantly reduce the feature set while maintaining or improving the classification accuracy using a decision tree classifier. Intrusion detection benchmark dataset, NSL-KDD, consisting of 41 features is used to evaluate the efficiency of our proposed method in Waikato environment for knowledge analysis (Weka).

The rest of the paper is organised as follows, related work is presented in Section 2 while the proposed EMFFS method is described in Section 3. In Section 4, the classification algorithm and benchmark dataset is presented while in Section 5, a detailed experimental result is discussed. Section 6 finally concludes the paper.

## 2. Related Work

The performance of a classification problem depends on the relevance of the selected attributes with regards to its class. Feature selection methods have been applied in classification problems to select a reduced feature subset from the original set to achieve a faster and more accurate classification. Similar to many data mining and machine learning techniques, two key factors are involved in building an optimum classifier; feature and model selection [9]. Selecting the right feature can be quite a challenging task, however, several methods have been proposed to solve this and discard redundant, irrelevant and noisy features. In this section, we review works that have been proposed in the literature.

Wang and Gombault [10] propose a filter selection method using IG and Chi-squared to extract 9 most important features from the 41 in the network traffic. Bayesian Network and C 4.5 (a decision tree classifier) were used to detect DDoS attack in the network. Results obtained shows that the detection accuracy remains the same while the overall efficiency improved. Bolon-Canedo et al. [11] combined discretizers, filters and classifiers to improve the classification performance by significantly reducing the feature set. This is applied to both binary and multi-class classification problems using KDD Cup 99 benchmark dataset. A supervised inductive learning approach called group method for data handling (GMDH) has been proposed in [12] using two variant techniques; monolithic and ensemble-based. Filter feature selection methods using IG, Gain Ratio and GMDH were used to rank features during the pre-processing phase. Lin et al. [13] propose an anomaly intrusion detection that detects new attacks using support vector machine (SVM), decision tree (DT) and simulated annealing (SA). The best features were selected from the KDD'99 dataset using SVM and SA to improve the classification accuracy of DT and SVM, to detect new attacks. Li et al. [16] propose a gradual feature removal method that process dataset prior to combining cluster method, ant colony algorithm and SVM to classify network traffic as either normal or anomaly. Sindhu et al. [14] propose a wrapper method for feature selection to remove irrelevant instances from a feature set to achieve higher detection accuracy using neuro tree. A feature selection approach has also been proposed in [17] using Bayesian Network. NSL-KDD dataset was used to evaluate the selected features and the result shows that the selected feature decreases attack detection time and improved the classification accuracy as well as the true positive rates. A recent work by Bhattacharya et al. [15] propose a multi-measure multi-weight ranking approach that identifies important network features by combining wrapper, filter and clustering methods to assign multiple weights to each feature.

Rough set feature selection approach has proven to be an efficient mathematical tool based on upper and lower approximation. It presents equal classification capability with the minimal subset. Olusola et al. [18] propose a rough set based feature selection method that selects important features from input data using KDD '99 dataset. Sengupta et al. [19] designed an online intrusion detection system (IDS) using both rough set theory and Q-learning algorithm to achieve a maximum classification algorithm to classify a data as either normal or anomaly using NSL-KDD network traffic data. A fast attribute reduction algorithm based on rough set theory was proposed in [36]. The algorithm identifies important features and discards independent and redundant attributes to achieve an effective classification performance.

Analyzing the reviewed work shows three general trends in feature selection irrespective of the method used. First, methods proposed search and identify correlated features in the dataset in order to remove the redundant features, some other methods identify unique features that contain important information about different output classes in the data and discards the ones with little or no information. Lastly, some features have been identified to be strong as a group but weak individually. In filter feature selection approach, features are ranked independently according to their strength in predicting the output class. Common filter methods present different ranking algorithms, therefore, we propose an EMFFS method that combines the output of IG, Gain Ratio, Chi-squared and ReliefF to find common features in the one-third split of the ranked features using NSL-KDD benchmark dataset in Weka environment. We, therefore, reduce the features from 41 to 13 and use J.4.8, a version of C4.5 decision tree classification algorithm to classify data as either normal or anomaly.

## 3. Ensemble-based Multi-Filter Feature Selection Method

The filter feature selection method is a pre-processing phase towards selecting important features from a dataset and is independent of the classification algorithm. Filter methods rely on statistical intrinsic test over an original training dataset and uses a feature ranking scheme as the main criteria for feature selection by ordering. Features are scored and a pre-determined threshold is used to remove features below the threshold. Due to its simplicity, it has been widely used for practical applications (i.e. cloud computing) involving a huge amount of data. In this section, we describe our proposed ensemble-based multi-filter feature selection method that combines the output of four filter selection methods i.e. IG, Gain Ratio, chi-squared and reliefF to harness their combined strength to select 13 common features among them.

A. Information Gain

One of the filter feature selection methods used in determining relevant attributes from a set of features is IG. IG works by reducing the uncertainty associated with identifying the class attribute when the value of the feature is unknown [21] and it is based on information theory used in ranking and selecting top features to reduce the feature size before the start of the learning process. The entropy value of the distribution is measured to determine the uncertainty of each feature before ranking it according to their relevance in determining different classes [20]. The uncertainty is determined by the entropy of the distribution, sample entropy or estimated model entropy of the dataset. The entropy of variable $X$ [22] can be defined as;

$$H(X) = -\sum_{i} P(x_i) log_2(P(x_i)) \qquad (1)$$

Let $P(x_i)$ denote the values of prior probabilities of $X$. The entropy of $X$ after observing value of another variable $Y$ is defined as,

$$H(X/Y) = -\sum_{j} P(y_j) \sum_{i} P(x_i|y_j) log_2\left(P(x_i|y_j)\right) \qquad (2)$$

In equation 2, $P(x_i|y_j)$ is the posterior probabilities of $X$ given the values of $Y$. The information gain is defined as the amount by which the entropy of $X$ decreases to reflect an additional information about $X$ provided by $Y$ and is defined as;

$$IG(X/Y) = H(X) - H(X|Y). \qquad (3)$$

Based on this measure, it is clear that feature $Y$ is said to be more correlated to feature $X$ than to feature Z, if $IG(X/Y) > IG(Z/Y)$. The feature ranking can therefore be calculated using equation 3. This ranking will be used to select the most important features.

B. Gain Ratio

The gain ratio was introduced to improve the bias of IG towards features with large diversity value [12]. When data are evenly spread, gain ratio exhibits a high value while it gives a small value when all data belongs to only one branch of the attribute. It uses both the number and size of branches to determine an attribute and corrects IG by considering intrinsic information [23]. The intrinsic information of a given feature can be determined by the entropy distribution of the instance value. Gain ratio of a given feature $x$ and a feature value $y$ can be calculated [23] using equation 4 and 5 below

$$Gain\ Ratio\ (y,x) = \frac{Information\ Gain(y,x)}{Intrinsic\ Value(x)} \qquad (4)$$

Where,

$$\text{Intrinsic Value }(x) = -\sum \frac{|S_i|}{|S|} * Log_2 \frac{|S_i|}{|S|} \qquad (5)$$

Note that $|S|$ is the number of possible values feature $x$ can take, while $|S_i|$ is the number of actual values of feature $x$. In our work, we selected 14 features, representing one-third split of the ranked features using NSL-KDD benchmark dataset. These 14 features represents the highest ranked feature using Gain Ratio.

C. Chi-squared

The chi-squared ($\chi^2$) statistic is used to test the independence of two variables by computing a score to measure the extent of independence of these two variables. In feature selection, $\chi^2$ measures the independence of features with respect to the class. The initial assumption of $\chi^2$ is that the feature and the class are independent before computing a score [24]. A score with large value indicates a high dependent relationship exists. Chi-squared [25] can be defined as:

$$\chi^2(r, c_i) = \frac{N[P(r,c_i)P(\bar{r},\bar{c}_i) - P(r,\bar{c}_i)P(\bar{r},c_i)]^2}{P(r)P(\bar{r})P(c_i)P(\bar{c}_i)} \qquad (6)$$

Where N denotes the entire dataset and r indicates the presence of the feature ($\bar{r}$ its absence), $c_i$ refers to the class. $P(r, c_i)$ is the probability that feature $r$ occurs in class $c_i$ and $P(\bar{r}, c_i)$ is the probability that the feature r does not occur in class $c_i$. Also, $P(r, \bar{c}_i)$ and $P(\bar{r}, \bar{c}_i)$ are the probabilities that the features does or does not occur in a class that is not labelled $c_i$ and so on. $P(r)$ is the probability that the feature appears in the dataset while $P(\bar{r})$ is the probability that the feature does not appear in the dataset. $P(c_i) \text{ and } P(\bar{c}_i)$ is the probability that a dataset is labelled to class $c_i$ or not.

D. ReliefF

ReliefF feature selection method uses continuous sampling to evaluate the worth of a feature to distinguish between the nearest hit and nearest miss (nearest neighbour from the same class and from a different class) [26]. The attribute evaluator is used to append weight to each feature according to its ability to distinguish the different classes. A user-defined threshold is determined and weight of features that exceeds this threshold are selected as important features [24]. ReliefF evolved from the original Relief algorithm [27] and was developed to improve its limitations. Among the key features of ReliefF are its ability to deal with the multiclass problem and its robustness and capability to deal with noisy and incomplete data. A key advantage of ReliefF over other filter methods is that it has a low bias and can be applied in all situations.

3.1 EMFFS execution process

Our proposed EMFFS method uses the output of the one-third split of ranked features of the filter methods described above. EMFFS is a pre-processing phase prior to learning where individual filter methods are used for the initial selection process. IG, gain-ratio, chi-square and reliefF filter methods are used to rank the feature set of the original dataset to create a mutually exclusive subset before selecting one-third split of the ranked features (i.e. 14 features). These features are considered as the most important feature with respect to each filter method.

The resulting output of the EMFFS is determined by combining the output of each filter method and using a simple majority vote to determine the final selected feature. A threshold is determined to identify the frequently occurring features among the four filter methods and set at 3 (i.e. $T=3$). After combining all the selected feature sets, a counter is used to determine common features with respect to the threshold set. Features that meets the threshold criteria are selected and used as the final feature set for classification. Figure 1, shows the proposed EMFFS method.

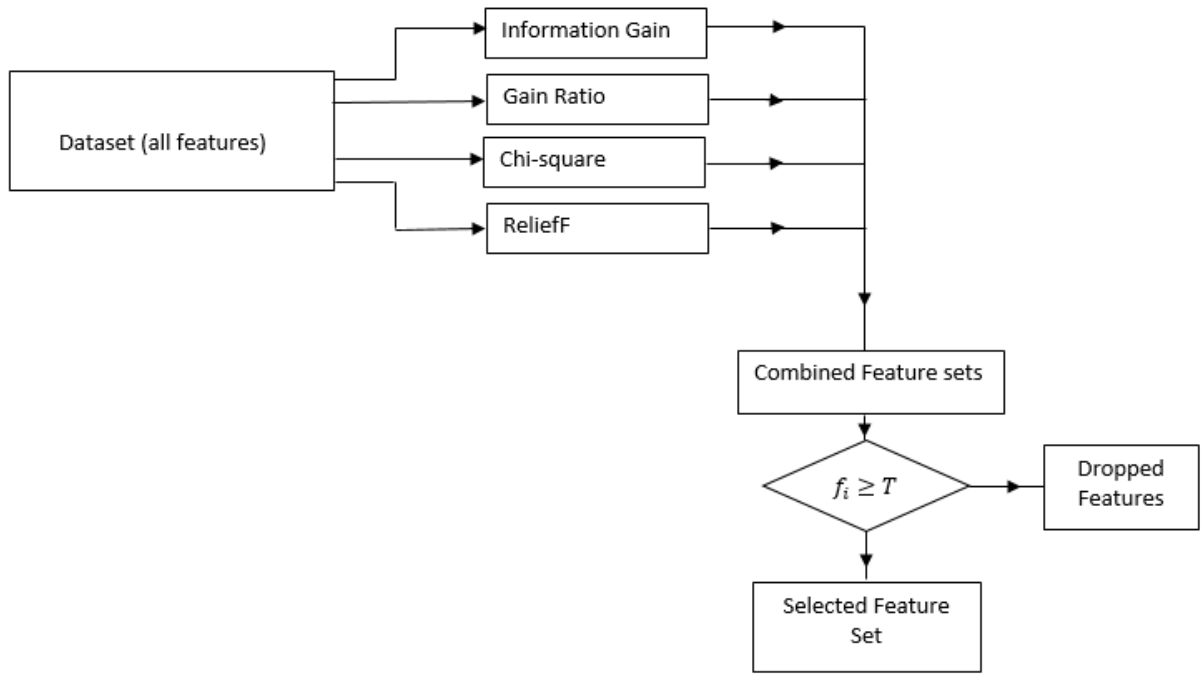

Figure 1 Ensemble-based multi-filter feature selection method

The EMFFS method is constructed through the algorithms presented below;

**Algorithm 3.1.1** (Filter feature ranking methods)

Step 1: Let $X_i$ be the feature set in the NSL-KDD dataset, where $X_i = \{X_1, X_2, X_3 \ldots \ldots \ldots \ldots, X_{41}\}$ and $C_i$ represents the class (i.e. normal or anomaly), where $C_i = \{C_1, C_2\}$.

Step 2: For each filter method, rank and sort the features $X_i$ according to its importance in determining the output class $C_i$.

Step 3: Select one-third split of each filter selection method's output $X_i'$.

**Algorithm 3.1.2** (Combine output features)

Step 1: Combine selected output features $X_i'$ of each filter method.

Step 2: Determine the feature count threshold $T$.

Step 3: Compute the feature occurrence rate among the filter methods.

**Algorithm 3.1.3** (Ensemble selection)

Step 1: Choose intercepts of common features from 3.1.2

Step 2: If the feature count is less than the threshold, drop the feature otherwise select the feature.

Step 3: Repeat step 2 for all the features in the one-third split subset.

## 4. Classification algorithm and dataset

Decision tree classification algorithm is a popular data mining classifier for prediction due to the ease of understating the interaction between variables. It is based on a greedy algorithm that uses a divide-and-conquer strategy to recursively construct a decision tree [28]. The tree is made up of the root node, internal nodes, branches and leaves, which represents a rule used in categorizing data according to its attributes. Decision tree uses supervised dataset with root node being the first attribute with the test condition to split each input towards individual internal node, in line with the characteristics of the data record [27]. The root node has the highest information gain, the preceding node with the next highest information gain is selected as the test for the next node. This process persists until the entire attributes are compared or when all the samples belong to the same class with no remaining attribute to which the samples can be further partitioned [29].

A branch connects two nodes together and can also connect a node and a leaf. Each node is made up of branches labelled as the possible value of attributes in the parent node [13]. The leaves are labelled as the decision value of classification.

Consider a case selected at random from a set S of cases which belongs to class $C_i$. The probability that an arbitrary sample belongs to class $C_i$ can be determined as follows [29]:

$$P_i = \frac{freq\ (C_i, S)}{|S|} \qquad (7)$$

Where $|S|$ is the number of samples in the set $S$. Therefore the information it convey can be represented by $-log_2 P_i$ bits. Now, suppose the probability distribution is given as $P = \{P_1, P_2, P_3 \ldots \ldots \ldots, P_n\}$, therefore, the information carried by the distribution, that is entropy of $P$, can be expressed as:

$$Info\ (P) = \sum_{i=1}^{n} -P_i\ log_2 P_i \qquad (8)$$

Partitioning a set of $K$ samples, based on the value of a non-categorical attribute X, into sets $K_1, K_2, K_3 \ldots \ldots \ldots, K_m$, the information required to determine the class of an element of $K$ is the weighted average of the information needed to identify the class of an element $K_i$. The weighted average of Info ($K_i$) can be determined by;

$$Info\,(X,K) = \sum_{i=1}^{m} \frac{|K_i|}{K} \times Info(K_i) \qquad (9)$$

The information gain, $Gain\,(X,K)$, can therefore be calculated as follows;

$$Gain\,(X,K) = Info\,(K) - Info\,(X,K) \qquad (10)$$

Equation 10 represent the difference between the information needed to identify an element of $K$ and an information needed to identify an element of $K$ after the value of attribute $X$ has been determined. Therefore this is the information gain due to attribute $X$.

There are different algorithms for implementing decision tree; C5.0 and its earlier version C4.5 has been described in [30], however, for our work, we will use J48, a version of C4.5 as our classifier.

4.1 Benchmark Datasets

NSL-KDD dataset, an improved version of KDDCUP'99 widely deployed in the literature for intrusion detection has been used to test our proposed algorithm. NSL-KDD is a labelled benchmark dataset developed from KDDCUP'99 to improve its flaws. Researchers have identified several issues that characterise KDDCUP'99, prominent among these are the presence of huge redundant records (which allows learning algorithm to be biased towards frequently occurring records) and its high complexity [31]. NSL-KDD is used for evaluating network intrusion systems and is made up of selected records from the initial KDDCUP'99. This presents a reduced dataset size that makes the evaluation of different research works consistent and validation of learning algorithm complete, easy and affordable. NSL-KDD is made up of 41 features and labelled as either attack or normal (see table 1). These features are categorized into four groups; basic features, content features, time-based traffic features and connection-based traffic features [15]. NSL-KDD has been divided into training and testing datasets. The training set is made up of 21 attack types while an additional 17 novel attack types are used for the test set [17]. The attacks are grouped into four categories: DoS, Probe, U2R and R2L. While the distribution of the training dataset consists of 67343 normal

(53.46%), 45927 DoS (36.46%), 11656 Probe (9.25%), 995 R2L (0.79%) and 52 (0.04%) U2R; the testing dataset on the other hand contains 9711 normal (43.08%), 7456 DoS (33.08%), 2421 probe (10.74%), 2756 R2L (12.22%) and 200 U2R (0.89%).

From the attack distribution, DoS constitutes around 78.3% of the total attack therefore in this work, we use 20% of the records in NSL-KDD train+ as our denial of service training set that has been labelled as either attack or normal. We apply 10-fold cross-validation for both training and testing purpose. Table 1 below shows the NSL-KDD feature dataset.

Table I: NSL-KDD dataset features

| # | Data features | # | Data features | # | Data features | # | Data features |
|---|---|---|---|---|---|---|---|
| 1 | Duration | 12 | Logged_in | 23 | Count | 34 | Dst_host_same_srv_rate |
| 2 | Protocol_type | 13 | Num_compromised | 24 | Srv_count | 35 | Dst_host_diff_srv_rate |
| 3 | Service | 14 | Root_shell | 25 | Serror_rate | 36 | Dst_host_same_src_port_rate |
| 4 | Flag | 15 | Su_attempted | 26 | Srv_serror_rate | 37 | Dst_host_srv_diff_host_rate |
| 5 | Src_bytes | 16 | Num_root | 27 | Rerror_rate | 38 | Dst_host_serror_rate |
| 6 | Dst_bytes | 17 | Num_file_creations | 28 | Srv_rerror_rate | 39 | Dst_host_srv_serror_rate |
| 7 | Land | 18 | Num_shells | 29 | Same_srv_rate | 40 | Dst_host_rerror_rate |
| 8 | Wrong_fragment | 19 | Num_access_files | 30 | Diff_srv_rate | 41 | Dst_host_srv_rerror_rate |
| 9 | Urgent | 20 | Num_outbound_cmds | 31 | Srv_diff_host_rate | | |
| 10 | Hot | 21 | Is_host_login | 32 | Dst_host_count | | |
| 11 | Num_failed_logins | 22 | Is_guest_login | 33 | Dst_host_srv_count | | |

## 5. Experimental results

In this paper, we deployed our proposed EMFFS method to pre-process the dataset to select the most important features for decision tree classification algorithm that classifies data as either attack or normal in cloud computing. All our experimental analysis has been carried out in Weka software [34] that contains a collection of machine learning algorithms for data mining tasks. The parameters for classification in all the experiments are set to the default values in Weka.

We use NSL-KDD dataset to evaluate the performance of our EMFFS method and decision tree classifier using 10-fold cross-validation. In the 10-fold cross-validation, data is divided into 10 folds of equal sizes before performing 10 iterations of training and validation. Within each iteration, a different fold of the data is used for validation while the remaining nine folds are used for learning. All experiments are performed on a 64-bit Windows 8.1 operating system with 6 GB of RAM and Intel core i5-4210U CPU.

5.1 Pre-processing Dataset

During the pre-processing phase, feature selection is performed to determine the most important features of NLS-KDD dataset, by ranking them, using different filter methods. Fourteen (14) most important features of the filter methods are determined by presenting one-third split of the ranked features. (-See table II)

Table II. Feature selection using filter methods

| Filter method | Feature selected |
|---|---|
| Info Gain | **5**,**3**,**6**,**4**,**30**,**29**,**33**,**34**,35,**38**,**12**,**39**,**25**,**23** |
| Gain Ratio | **12**,26,**4**,**25**,**39**,**6**,**30**,**38**,**5**,**29**,**3**,37,**34**,**33** |
| Chi- Squared | **5**,**3**,**6**,**4**,**29**,**30**,**33**,**34**,35,**12**,**23**,**38**,**25**,**39** |
| Relief-F | **3**,**29**,**4**,32,38,**33**,**39**,**12**,36,**23**,26,**34**,40,31 |

After applying algorithm 3.1.2 to the output of each of the four filter selection method, we search for feature intercept and set the minimum threshold to 3. From Table II, it is observed that even though each filter uses different ranking techniques, some features are common across different filter methods. Using simple majority vote, features 4, 29, 34, 12, 39, 3, 5, 6,30,33,38, 25 and 23 (indicated in bold) appear across more than three filter methods; this shows how important these features are to the output class (-See table III).

Table III. Multi-filter features selection method

| Filter method | Feature selected |
|---|---|
| Multi-filter | 3,4,29,33,34,12,39,5,30,38,25,23,6 |

Table III shows the 13 selected features out of the one-third split of the most important features of NSL-KDD dataset using EMFFS method. This will be used as the input features for training the decision tree classification algorithm, J48, in Weka.

5.2 Performance measures

The performance of a classifier can be determined by using different metrics. Determining the accuracy usually involves the measure of True Positive (TP), True Negative (TN), False Positive (FP) and False Negative (FN). TP is the of attack classified correctly while TN is the percentage of normal test sample classified correctly. FP is the amount of attack detected when it is indeed normal (false alarm) and FN is the misclassification of a test sample as normal when it is actually an attack.

Recently developed systems for attack detection requires high detection rate and low false alarm, therefore in this work, we compare the accuracy, detection rate and false alarm rate of our proposed EMFFS method with each filter method and the full dataset feature using J48 classification algorithm. Furthermore, we compare the time to build the classification model, which is the duration of classifier learning after applying each feature selection method.

Table IV presents the results of the performance measure of the J48 classifier using the full dataset with 41 features, one-third split of filter methods with 14 features and our proposed EMFFS method with 13 features.

Table IV Performance measure

| Filter method | No of features | Accuracy | Detection rate | False alarm rate | Time |
|---|---|---|---|---|---|
| Full set | 41 | 99.56% | 99.49% | **0.38%** | 2.75 Sec |
| Info Gain | 14 | 99.66% | 99.74% | 0.41% | 0.83 Sec |
| Gain Ratio | 14 | 99.60% | 99.68% | 0.47% | 1.12 Sec |
| Chi-squared | 14 | 99.66% | 99.74% | 0.41% | 0.92 Sec |
| ReliefF | 14 | 99.08% | 99.02% | 0.87% | 0.93 Sec |
| Multi-filter | 13 | **99.67%** | **99.76%** | 0.42% | **0.78 Sec** |

Classification accuracy

Classification accuracy is the percentage of correctly defined data from the total set represented by the situation of TP and TN. The accuracy of the classifier can be determined by;

$$Accuracy = \frac{TP+TN}{TP+TN+FP+FN} \times 100\%$$

Figure 2 shows the classification accuracy across different filter feature selection methods and EMFFS method. Our proposed method presents a slight improvement in performance.

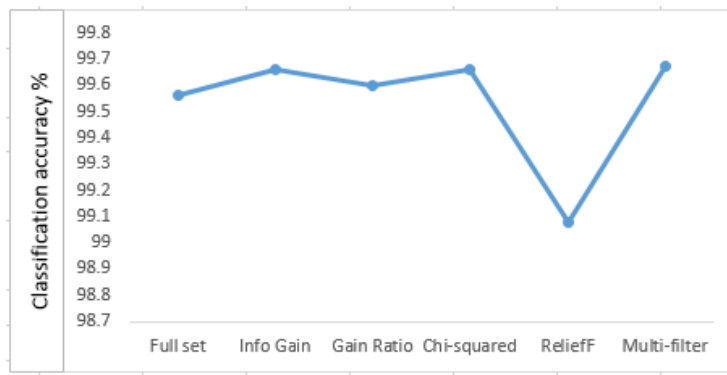

Figure 2 Classification accuracy for filter methods

Detection rate

Detection rate can be determined based on the confusion matrix. It is calculated as follows; $Detection\ rate = \frac{TP}{TP+FN} \times 100\%$.

Figure 3 shows the performance of EMFFS method in comparsion to other filter feature selection methods. The results presented show that our method with 13 selected features has a slight improvement in detection rate when compared with other filter methods.

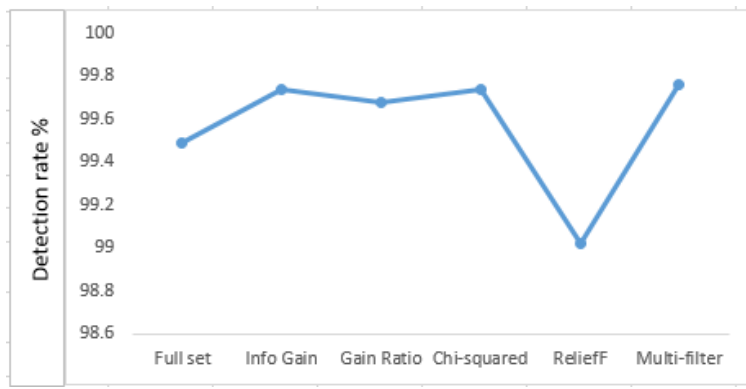

Figure 3 Detection rate for filter methods

False alarm rate

False alarm is the amount of normal data that has been falsely classified as an attack, this can be determined by $False\ alarm\ rate = \frac{FP}{FP+TN} \times 100\%$.

Figure 4 shows the false alarm rate of the full feature set and different filter feature selection methods. ReliefF produces the highest false alarm rate while the full feature set having the best performance with 0.38%. Our method presents a false alarm rate of 0.42%.

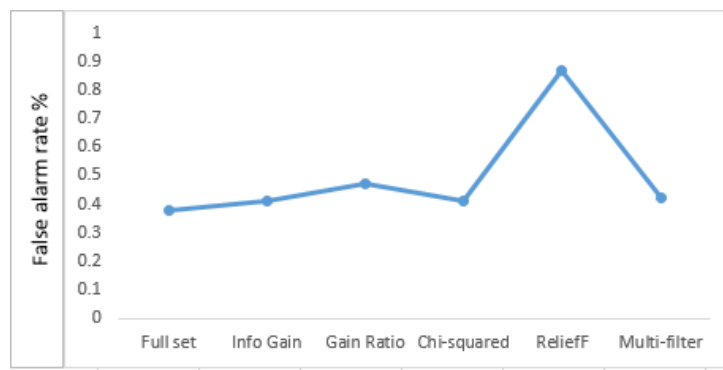

Figure 4 False alarm rate for filter methods

Time to build model

Figure 5 presents the time to build model across different filter selection methods and the full feature set. Our proposed method presents the best time with 0.78 sec when compared with other filter selection methods. The full feature set presents the worst learning time with 2.75 sec. This is due to the number of features the classifier have to process.

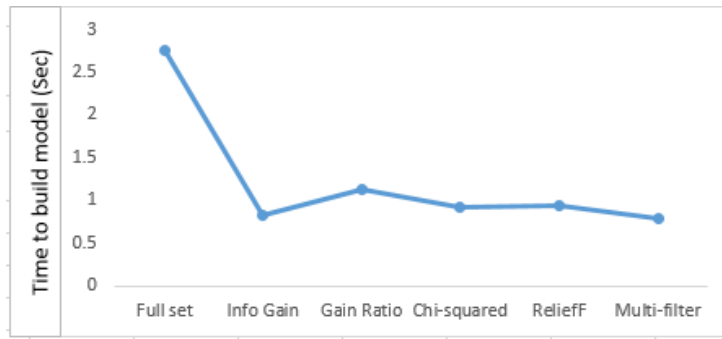

Figure 5 Time to build model for filter methods

5.3 Discussion

The need for effective real-time classification of DDoS attack in cloud computing increases the complexity of proposed detection techniques. Filter methods for feature selection have proved to be a crucial path towards building a light weight detection system, which involves identifying important features. In our proposed EMFFS method, we have selected 13 features out of available 41 features by first presenting the output of one-third split using four filter methods. We determined a threshold and used a counter to select important features by simple majority voting. We compare our EMFFS method with other filter methods with 14 features and the full set consisting of 41 features using J48 decision tree classifier. Our method with 13 features presents an improvement in classification accuracy and detection rate. This implies that the original dataset contains some level of redundant feature that has little or no contribution towards identifying a particular class. For the time taken to build the model, our proposed method presents the best time when compared with individual filter selection methods and the full feature set. This makes our ensemble-based multi-filter feature selection method efficient with less complexity.

We have also compared the performance of our proposed method, EMFFS, with methods proposed in the literature by considering numbers of feature selected, classification accuracy and time to build model as shown in Table V. We observed that using 13 most important features with decision tree classifier, our method produced the best classification accuracy and a more efficient better learning time, in comparision to the other schemes presented in Table V.

Table V Performance measure

| Approach | Classifier | No of features | Accuracy (%) | Time to build model (sec) |
|---|---|---|---|---|
| CFS [34] | C4.5 | NA | 99.13 | NA[1] |
| CFS, CONS & INTERACT[9] | HNB_PKI_INT | 7 | 93.72 | NA[1] |
| Gradual feature removal [5] | Cluster methods, ant colony algorithm & SVM | 19 | 98.62 | NA[1] |
| CSE & CFS [33] | GA | 32 | 78 | NA[1] |
| Linear correlation-based [35] | C.45 | 17 | 99.1 | 12.02 |
| Our method | J48 | 13 | **99.67** | **0.78** |

NA[1]: Not available

## 6. Conclusion

One of the notable challenges faced by current network intrusion systems in cloud computing is the handling of massive internet traffic during DDoS attack. Feature selection methods have been used to pre-process dataset prior to attack classification in cloud computing. This work has presented an ensemble-based multi-filter feature selection method that combines the output of one-third split of ranked important features of information gain, gain ratio, chi-squared and reliefF. The resulting output of the EMFFS is determined by combining the output of each filter method and using a set threshold to determine the final feature using a simple majority vote. Performance evaluation with NSL-KDD dataset shows that EMFFS method with 13 features has a better performance than other filter methods using J48 classifier and other proposed feature selection methods.

In future work, we plan to extend our work to include other classification algorithms and evaluate using other publicly available labelled datasets.